# Modulating Thermal Conductivity via Targeted Phonon Excitation

Xiao Wan[1#], Dongkai Pan[1#], Jing-Tao Lü[2], Sebastian Volz[3,4], Lifa Zhang[5], Qing Hao[6], Yangjun Qin[1], Zhicheng Zong[1], Nuo Yang[1*]

1. School of Energy and Power Engineering, Huazhong University of Science and Technology, Wuhan 430074, China.
2. School of Physics and Wuhan National High Magnetic Field Center, Huazhong University of Science and Technology, Wuhan 430074, China.
3. LIMMS, CNRS-IIS UMI 2820, The University of Tokyo, Tokyo 153-8505, Japan.
4. Institute of Industrial Science, The University of Tokyo, Tokyo 153-8505, Japan.
5. NNU-SULI Thermal Energy Research Center and Center for Quantum Transport and Thermal Energy Science (CQTES), School of Physics and Technology, Nanjing Normal University, Nanjing 210023, China.
6. Department of Aerospace and Mechanical Engineering, University of Arizona, Tucson, AZ, 85721-0119, USA.

\# X. W. and D. P. contributed equally to this work.
*Corresponding email: nuo@hust.edu.cn (N.Y)

# Abstract

Thermal conductivity is a critical material property in numerous applications, such as those related to thermoelectric devices and heat dissipation. Effectively modulating thermal conductivity has become a great concern in the field of heat conduction. In this study, a quantum strategy is proposed to modulate thermal conductivity by exciting targeted phonons. The results show that the thermal conductivity of graphene can be tailored in the range of 1559 W/m-K (49%) to 4093 W/m-K (128%), compared with the intrinsic value of 3189 W/m-K. A similar trend is also observed for graphene nanoribbons. The results are obtained through both ab initio calculations and molecular dynamics simulations. This brand-new quantum strategy to modulate thermal conductivity paves a way for quantum heat conduction.

# Introduction

The modulation of thermal conductivity is of great importance in various applications, including thermal managements [1-3] and energy devices [4-6]. It means that thermal conductivity, as an intrinsic property of materials, can be enhanced or decreased by an external field. And then, the heat transfer characteristics of the material are altered. Achieving desirable performances requires a deep understanding of phonon scattering mechanisms in different heat transfer regimes, which is essential to modulate the thermal conductivity of materials [7-16]. However, the complex nature of phonon transport has made the modulation of thermal conductivity a long-standing challenge in physics and material science.

Recent developments in the field of heat conduction have led to a better understanding of the scattering dynamics of heat carriers at the nanoscale. Heat conduction in dielectrics can be understood as the propagation of phonons and their scatterings such as phonon-phonon [17-20], impurity [21-23] and boundary scattering [9,24,25]. Phonon-phonon scattering has been exploited to produce weaker couplings [26-28] and to highlight hydrodynamic phonon transport [11,29] in nanostructures. Impurity scattering is highly frequency dependent and also closely related to normal processes [30], which can redistribute phonon frequencies and control phonon transport by nano-engineering [1,15,21,31,32]. Due to size confinement, phonon transport is largely affected by the boundary scattering [9], resulting in the size dependence of thermal conductivity and leading to an invalid Fourier's law [33-35].

As the understanding of phonon scattering mechanisms has advanced, a variety of new strategies to modulate the thermal conductivity have been developed, motivated by a great demand for thermal management. An enhanced thermal conductivity can be achieved by minimizing phonon–phonon scattering phase space and phonon–impurity scattering in bulk materials [2,14,18-20,36-42]. Conversely, to reduce thermal conductivity, the strategies that increase phonon scattering have been explored [43-46], such as intrinsically increasing anharmonicity or crystal complexity, or extrinsically introducing disorder, defects, boundaries, interfaces and nanoparticles. Besides, the wave nature of phonon [47] can also be leveraged to modulate the thermal conductivity, as seen in the development of nanophononic crystals [48-50] and the raise of tuning phonon coherence [51-53] and localization [54-56]. In recent works, the external fields have been implemented to change the morphology of structures, thereby regulating thermal conductivity [37,41,57-59]. Nevertheless, while these strategies mentioned above have shown promise, they currently cannot modulate the thermal conductivity with quantum precision or controllability.

Recently, there has been a growing interest in the targeted phonon mode excitation using the terahertz optical pulses or tensile strain to modulate optical and electrical properties. The methods offer several advantages such as in-situ, flexibility, quick response, and directness, without requiring any structural modification. For instance, terahertz excitation pulses have been used to directly excite optical phonons in $MAPbI_3$, resulting in significant perturbations in the electron relaxation dynamics [60]. Similarly, selectively exciting vibrational modes of the molecules has been shown to modulate the performance of an organic optoelectronic system [61]. Furthermore, the non-equilibrium carrier-phonon dynamics in photovoltaic systems have been discussed in details for a few perovskites. These discussions have revealed the quantum emission of longitudinal optical (LO) phonons, the decay of the optical phonon to acoustic phonons, and other relaxation processes when abundant carriers are injected [62-64]. Non-equilibrium between optical and acoustic phonons has also been observed in photoexcited graphene and $MoS_2$ [65,66]. In black phosphorene, the excitation of out-of-plane acoustic phonons induced by tensile strain can provide a strong modulation of the electronic band structures, carrier lifetime and carrier mobility [67]. Additionally, targeted phonon excitation has been used to enhance ion diffusion [68] and induce structural phase transitions [69,70]. While these results demonstrate the feasibility of targeted phonon excitation, fewer studies have investigated the strategy's potential for modulating thermal conductivity.

Here, the strategy of quantum excitation of phonons is proposed to modulate thermal conductivity of dielectric materials, where phonons dominate in the heat conduction. This strategy involves exciting targeted phonons to increase or decrease phonon scattering, thus achieving the desired thermal conductivities. The effectiveness and capability of this strategy are demonstrated using ab initio calculations [71] and molecular dynamics simulations [72]. Graphene and graphene nanoribbon are chosen as the model systems since the thermal conductivities of graphene and its derivatives have been extensively studied. The results indicate that the thermal conductivity can be modulated by exciting phonons in a quantum manner, without altering the structure.

## Strategy for Controlling Thermal Conductivity

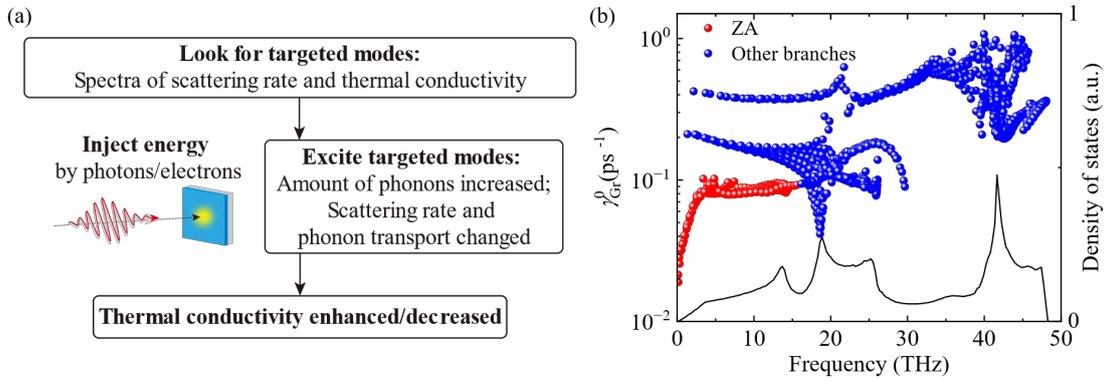

Fig. 1. Modulating thermal conductivity via targeted phonon excitation. (a) Schematic of modulation strategy. (b) The density of states (DOS) and intrinsic scattering rates $\gamma_{\text{Gr}}^0$ of graphene (Gr) by ab initio calculations for choosing targeted phonons. The superscript "0" represents the intrinsic value.

The strategy aims to modulate thermal conductivity by exciting targeted phonon modes in a quantum manner, as illustrated in Fig. 1(a). Firstly, the dominant modes for heat transport are identified utilizing first principles and molecular dynamics simulations. Then, the energy of these modes is artificially increased, allowing for the quantum excitation of dominant phonons with large contributions to thermal conductivity and weak coupling with other phonons. This results in a significant enhancement of thermal conductivity. On the other hand, if phonons with high scattering rates and relatively low contributions to thermal conductivity are excited, scattering processes are promoted, leading to a decrease in thermal conductivity. Phonon DOS determines the number of excited modes around a specific frequency, thus indicates the effectiveness of activating those modes. It is worth noting that thermal conductivity can be modulated over a wide range by exciting only a few dominant phonons, without introducing other scattering mechanisms.

The method is basically relying on the nature of phonon scattering. Heat conduction in solids is directly related to phonon scattering, where an individual mode can participate in various scattering processes. By identifying the characteristic time of phonon relaxation time or lifetime, $\tau$, the thermal conductivity can be generally written as an integration as [9,10,24]:

$$\kappa = \frac{1}{3}\int C(\omega)v^2(\omega)\tau(\omega)d\omega \qquad (1)$$

where parameter $1/3$ is related to the dimension of the system, $C(\omega)$ refers to the spectral volumetric specific heat, $v$ to the phonon group velocity and $\omega$ is the frequency. In general, the diverse scatterings can be incorporated into the Matthiessen's rule [9,24]

$$\frac{1}{\tau} = \frac{1}{\tau_{ph-ph}} + \frac{1}{\tau_{im}} + \frac{1}{\tau_b} \qquad (2)$$

where $\tau_{ph-ph}$, $\tau_{im}$, and $\tau_b$ are the relaxation times due to phonon-phonon, impurity, and boundary scattering processes, respectively. The relaxation time of phonon-phonon scattering is calculated using Fermi's golden rule [73], while impurity scattering can be attained based on Klemens' derivation [9]. In addition, crystal boundary scattering is determined by diffuse boundary absorption/emission, which depends on the Casimir length. It is important to note that impurity and boundary scattering aren't included in the calculation of this work.

By increasing the energy of targeted phonons, the scattering rate can be changed significantly, leading to the modulation on the thermal conductivity. To change the energy of mode $n$, the phonon occupation numbers are modified in ab initio calculations according to the following formulation:

$$E'_n = \hbar\omega_n \frac{N}{e^{\frac{\hbar\omega_n}{k_B T}} - 1} \tag{3}$$

where $E'_n$ corresponds to the energy of mode $n$ after the modification, $\hbar$ denotes the reduced Planck constant, $k_B$ the Boltzmann's constant, $T$ is the absolute lattice temperature and $N$ represents the multiple of energy increase. To rescale the mode kinetic energy $E_{n,K}$ in molecular dynamic (MD) simulations, the atomic velocities are rescaled according to the formulation (The derivation is presented in Appendix C):

$$\mathbf{v}'_j = \mathbf{v}_j + \frac{1}{\sqrt{m_j}}(\sqrt{M} - 1)\dot{Q}_n(t)\mathbf{e}_{j,n} \tag{4}$$

where $\mathbf{v}'_j$ and $\mathbf{v}_j$ are the velocity of atom $j$ after and before the modification, respectively. The rescale factor $M$ is set to 10 in MD simulations. In addition, $m_j$ indicate the atomic mass, $\dot{Q}_n$ the normal mode velocity coordinate and $\mathbf{e}_{j,n}$ the eigenvector. Considering the time evolution of mode kinetic energy in the MD simulation (as shown in Fig. S4), the actual multiple of kinetic energy increase $N$ is much lower than $M$, and it can be defined as the relative time-averaged mode kinetic energy after and before the excitation $\langle E'_{n,K}\rangle/\langle E_{n,K}\rangle$. The effectiveness of the strategy is demonstrated in both graphene and graphene nanoribbon systems. However, it is worth noting that this approach could be applicable to other systems with weak-coupling phonons [26-28]. Ab initio calculations (Details in Appendix A) for graphene are performed using the Vienna ab initio simulation package (VASP) [74], and the phonon transport properties are obtained by solving the phonon Boltzmann transport equation with the special version of ShengBTE packages modified by Ruan et al. which has better convergence over two-dimensional materials [75,76], assisted by the PHONOPY package [77]. Additionally, nonequilibrium molecular dynamics (NEMD, Details in Appendix B) simulations for graphene nanoribbon are conducted using the LAMMPS package [78].

To experimentally realize the proposed strategy, the key step is to excite the targeted phonon modes. This can be achieved by illuminating the sample with THz pulses using a tabletop light source, which are generated by optical rectification using the tilted-pulse-intensity-front scheme. The intensity of the THz pulse can be controlled by a pair of wire grid polarizers operating in the THz frequency region, as described in Ref. [60]. Ideally, the frequency of the illuminated photons should be close to the THz magnitude, corresponding to the "targeted phonons" indicated in the thermal conductivity spectrum in Fig. 2a in the manuscript. Techniques such as THz time domain spectroscopy (THz-TDS) can be utilized to detect any enhanced signal caused by external stimuli and compare it to the baseline (no shining) performance.

## Results and Discussions

Firstly, the spectral contribution to intrinsic thermal conductivity of graphene $\kappa_{\text{Gr}}^0$ is calculated as a guidance to highlight the effectiveness of the modulation strategy, as shown in Fig. 2(a). An iterative solving method is utilized to accurately solve the Boltzmann transport equation [75]. The obtained value for $\kappa_{\text{Gr}}^0$ at 300 K is 3189 W m$^{-1}$ K$^{-1}$, which is comparable to previous works (1500−4000 W m$^{-1}$ K$^{-1}$) [79-83]. Fig. 2(a) suggests that the low frequency modes dominate the heat conduction in graphene particularly those below 5 THz and around 10 THz. The spectral $\kappa_{\text{Gr}}^0$ drops drastically in the higher frequency range, indicating a negligible contribution of these modes to heat conduction. This observation provides a prominent information for selecting the modes to be adjusted in energy.

Then, the energy is injected to excite more heat carriers into those dominant modes.

However, it can be observed from Fig. 2(b) that the results do not correspond perfectly to the tendency shown in Fig. 2(a), since the DOS and scattering rates are also essential factors in implementing the modification, as illustrated in Fig. 1(b). The DOS determines the number of modes that can be affected during excitation, while a high scattering rate means that the modes are strongly scattered, leading to a smaller contribution to heat conduction and a larger resistance of other modes. In graphene, ZA modes play a much more significant role in heat conduction due to the symmetry-based selection rule [82]. The reflection symmetry of its 2D structure excludes all the 3-phonon scatterings that involve odd numbers of flexural phonons, causing a profound reduction in the scattering rates of flexural modes, especially ZA modes. Thus, exciting those modes can generate more noticeable effects. As a result, the peak of $\kappa_{Gr}/\kappa_{Gr}^0$ located at around 1.7 THz, ensures a high enough DOS, a low scattering rate as well as a high spectral thermal conductivity. On the other hand, the trough is situated around 13.9 THz, close to the maximum frequency of the ZA phonon branch, ensuring a high enough DOS, a relatively high scattering rate, and small spectral thermal conductivity compared to other ZA modes. The scattering rates' changes after and before the excitation are also presented in Appendix A.

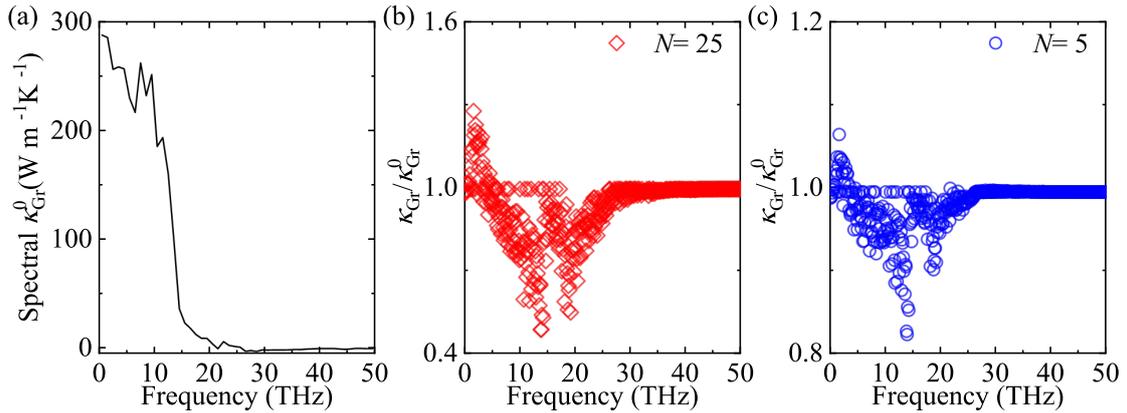

Fig. 2. Thermal conductivity modulation for graphene via targeted phonon excitation. (a) Spectral contributions to $\kappa_{Gr}^0$ as a function of frequency. (b, c) The relative overall thermal conductivity of graphene after and before the excitation $\kappa_{Gr}/\kappa_{Gr}^0$ as a function of the center frequency for targeted phonons. For instance, the first center frequency is 0.05 THz, and a few modes in 0−0.1 THz are excited with a twenty-five- or five-times

larger energy. The energy needed for the excitation is at most 20.8 mJ m$^{-2}$ ($N$=25) or 3.5 mJ m$^{-2}$ ($N$=5).

Molecular dynamics simulations are conducted on graphene nanoribbon to validate the feasibility of the modulation strategy, especially considering high-order scatterings. As shown in Fig. 3(a), the thermal conductivity contribution spectrum of graphene nanoribbon (24.9×21.6 Å$^2$) are extracted through NEMD simulations [84,85]. The intrinsic thermal conductivity of graphene nanoribbon $\kappa_{GNR}^0$ at 300 K is 59.0 W m$^{-1}$ K$^{-1}$, which is consistent with previous studies [49]. Due to the limitation in the simulation cell size, the possible excited phonons have smaller wavelengths and higher frequencies (above 3 THz). The thermal conductivity contribution spectrum is obtained by extracting time-dependent atomic motion data in the intermediate region where the temperature difference remains small (<0.3 K in all the cases) to ensure the validity of the lattice dynamic method. The excitation region is chosen to be all parts of the simulation cell except for the heat sink and fixed region, with the additional heat flux generated by the excitation being relatively small (2−5%) compared to the heat flux generated by the Langevin thermostat (Fig. S5 in Appendix B). The excitation is conducted every 100 fs with an equivalent input energy density of 0.21−0.48 mJ m$^{-2}$. The overall thermal conductivity refers to an effective quantity measuring the heat flux. It should also be noted that there are only a few modes in each 1 THz frequency range, which accounts for 2−7% of all modes of the Brillouin Zone. Moreover, modes around 10 THz make a significant contribution to the intrinsic thermal conductivity, as shown in Fig. 3(a). In particular, selecting two modes between 9 and 10 THz as targeted phonons (excited-A) results in a considerable increase in the overall thermal conductivity, with a 97.8% enhancement (from 59.0 to 116.7 W m$^{-1}$ K$^{-1}$) when those modes are excited with $N$=3.34 (3.34 times larger kinetic energy). Fig. 3(a) also shows the thermal conductivity contribution after the excitation. It can be concluded that the increase of $\kappa_{GNR}$ is mainly contributed from modes below 10 THz and modes in the 24−26 THz range, with the contribution of these modes not changing significantly compared to the original value. Furthermore, targeted phonons in other frequency range are also excited in a similar manner, as shown in Fig. 3(b). The excitation of these modes regulates the overall thermal conductivity by different ratios. The modes between 9 and 10 THz show the highest modulation ratio, owing to high contribution to $\kappa_{GNR}$, with the modes in the 25−26 THz range showing similar behavior. In general, the enhancement of $\kappa_{GNR}$ decreases with an increased frequency (below 15 THz). This can be attributed to higher phonon DOS in the high frequency range.

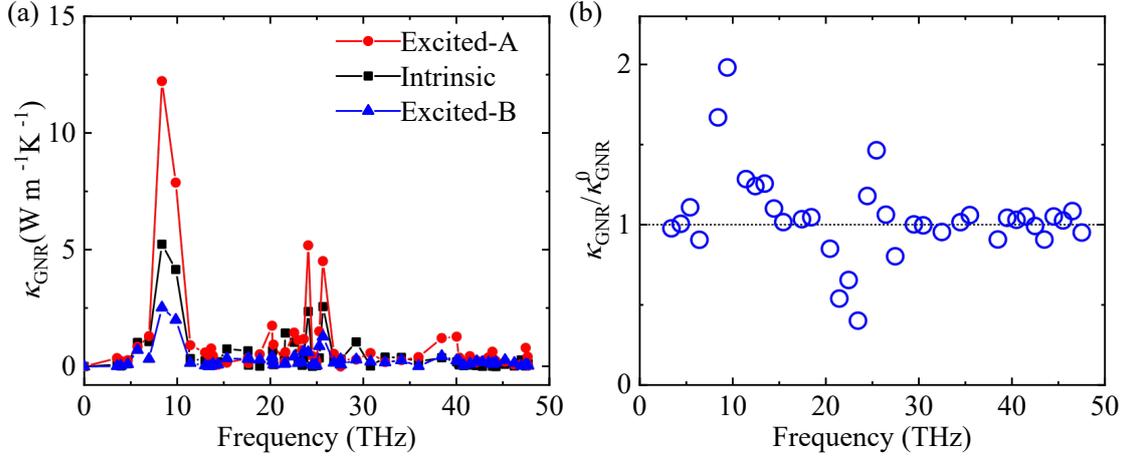

Fig. 3. Thermal conductivity modulation for graphene nanoribbon (GNR) via targeted excitation of phonons. (a) Thermal conductivity contribution spectrum by molecular dynamics simulations, "A" refers to the case where two modes, 9.84 THz, $\mathbf{k} = (-0.64, 0.96, 0)$ and 9.84 THz, $\mathbf{k} = (0.64, -0.96, 0)$, are excited with a 3.34 times larger kinetic energy (N=3.34), and "B" represents the case where four modes, 23.59 THz, $\mathbf{k} = (-0.64, 0.37, 0)$; 23.41 THz, $\mathbf{k} = (-0.64, 0.66, 0)$; 23.41 THz, $\mathbf{k} = (0.64, -0.66, 0)$; and 23.59 THz, $\mathbf{k} = (0.64, -0.37, 0)$, are excited with N=2.71. (b) Relative thermal conductivity of graphene nanoribbon after and before the excitation $\kappa_{GNR}/\kappa_{GNR}^0$ by exciting a few modes in each 1 THz frequency range utilizing molecular dynamics simulations.

The modulation results of NEMD simulations for decreasing $\kappa_{GNR}$ exhibit a similar tendency to the ab initio calculations. In Fig. 3(a), modes in the 15−24 THz range show low contribution to $\kappa_{GNR}$. When four modes between 23 and 24 THz are excited with a 2.71 times larger kinetic energy (excited-B, N=2.71), $\kappa_{GNR}$ decreases by 59.6% (from 59.0 to 23.8 W m$^{-1}$ K$^{-1}$). The comparison before and after the excitation is shown in Fig. 3(a). The contribution of modes to $\kappa_{GNR}$ decreases over a wide frequency range with a greater reduction below 10 THz due to the large original contribution values. Additionally, exciting the modes in the frequencies between 20 and 24 THz also reduce $\kappa_{GNR}$ (from 85.1% to 40.4%), as illustrated in Fig. 3(b). The reduction of $\kappa_{GNR}$ is significant and falls as the frequency of targeted phonons increases (below 24 THz). This trend can be related to high intrinsic scattering rates and low phonon DOS of modes around 23 THz compared to other modes.

## Conclusions

In this work, a new strategy for modulating thermal conductivity is proposed, which is realized by exciting targeted phonons in a quantum manner. The results demonstrate that this strategy can effectively modulate the thermal conductivity of graphene and graphene nanoribbons over a wide range compared to their intrinsic values. Ab initio calculations and NEMD simulations provide a detailed validation for the strategy. Firstly, the modes with top-contribution or strong-scattering are identified; and then, the energy of some of these modes is artificially increased.

It is worth noting that the feasibility of this strategy has a high possibility to be confirmed experimentally [60,61,70]. Because the quantum excitation of vibrational modes has already been used to modulate transport properties and induce structural phase transitions [61,67,68,70]. Furthermore, these findings provide a promising quantum approach to modulating thermal conductivity at the scale of phonon modes.

## Acknowledgements

This work is sponsored by the National Key Research and Development Project of China No. 2018YFE0127800. We are grateful to Lina Yang for useful discussions. The authors thank the National Supercomputing Center in Tianjin (NSCC-TJ) and the China Scientific Computing Grid (ScGrid) for providing assistance in computations.

## Declaration of Competing Interest

There are no conflicts of interest to declare.